\documentstyle{article}
\input boxedeps.tex
\SetRokickiEPSFSpecial  
\HideDisplacementBoxes
\def\d{\dagger}
\def\a{\alpha}
\def\b{\beta}

\def\be{\begin{equation}}
\def\eq{\end{equation}}
\def\Tr{{\rm \, Tr }}
\def\tsi{\tilde{\psi}}

\def\p{p^{+}}
\def\k{k^{+}}
\def\g{g^{2}_{N}}

\begin{document}

\mbox{}\hfill CERN-TH/97-198\\
\mbox{}\hfill hep-th/9708115\\
\vspace{30mm}
\begin{center}
{\LARGE  Adjoint $QCD_2$ and the Non-Abelian Schwinger Mechanism \\}
\vspace{30mm}

S. DALLEY {}\footnote{CERN Fellow, on leave from
Department of Applied Mathematics and Theoretical Physics, Cambridge
University, Silver Street, Cambridge CB3 9EW, England.}\\
\vspace{10mm}

{\em Theory Division, CERN \\
CH-1211 Geneva 23, Switzerland\\}

\vspace{30mm}

\end{center}

\begin{abstract}
Massless Majorana fermions in the adjoint representation of $SU(N_c)$
are expected to screen  gauge interactions in $1+1$ dimensions, analogous to a 
similar Higgs phenomena known for $1+1$-dimensional $U(1)$ gauge theory with 
massless fundamental fermions (Schwinger model). Using the light-cone 
formalism and large-$N_c$ limit, a 
non-abelian analogue of the Schwinger boson is shown to 
be responsible for the screening between heavy test charges.
This adjoint boson does not exist simply
as a physical state, but boundstates are
built entirely from this particle.

\end{abstract}

\newpage
\baselineskip .25in
\section{Introduction}

In a classic paper \cite{schwinger}, Schwinger showed that the photon of
two-dimensional QED acquires a propagating longitudinal component
of mass $e /\sqrt{\pi}$
by a dynamical version of the Higgs mechanism involving
massless electrons. This solvable model has been a source of much inspiration
in the development of gauge theories of particle physics. It is natural to
ask if there is a similarly tractable model involving non-abelian gauge theory.
In this paper we will show that by adding additional matter 
representations in the adjoint representation
to $QCD$ in $1+1$ dimensions ($QCD_2$), one has a very close
analogue of the Schwinger mechanism.
Such models are interesting because one may study 
exactly some of the non-perturbative effects of the 
zero transverse momentum modes of gluons or gluinos from $3+1$ dimensions. 
Even when exact results are not possible, the models are useful for testing
numerical algorithms which can then be applied in higher dimensions.

Recent interest in the problem of adjoint $QCD_2$ began with
the light-cone quantisation of the large-$N_c$ limit performed by
Klebanov and the author \cite{dalley1}. Numerical solution of the
light-cone Schrodinger equation for singlet boundstates of
adjoint quanta revealed  repeated `Regge
trajectories'
--- a kind of glueball analogue of the single meson
trajectory found by 't Hooft for large-$N_c$ $QCD_2$ with fundamental fermions
\cite{hoof}. In the low-lying spectrum these trajectories 
could be accurately classified by the number of adjoint quanta in a 
boundstate, but since
the number of these particles in not conserved, even at large $N_c$, it
was found that at higher mass this simple picture broke down for light
quanta. Further
numerical and analytic work \cite{kutasov1,klebanov1,klebanov2,dalley2} 
confirmed these conclusions and it was also suggested that generically 
the density of bound states rises exponentially, leading
to a Hagedorn transition \cite{kutasov1, kutasov3}. 

New insight into the problem involving massless
adjoint fermions came from an observation of
Kutasov and Schwimmer \cite{kutasov2}, who showed that for massless
fermions in two-dimensional gauge theory (not necessarily large $N_c$)
the massive physics is largely independent of
the representations present, provided they make up the same value of the
chiral anomaly in each of the two chiral sectors, so that the total
anomaly cancels.
This result is especially clear in the light-cone formalism,
where massless left moving fermions decouple (for a quantisation surface
$x^+ = (x^0 + x^1)/\sqrt{2} = {\rm const.}$) from the  Hilbert
space of massive boundstates. Physical results are thus insensitive
to most of the details of the left-moving  representations. Choosing instead
the quantisation surface $x^- = (x^0 - x^1)/\sqrt{2} = {\rm const.}$,
one arrives at the same conclusion for the right-moving fermions
and therefore the entire theory. One consequence of this universality
is that the massive physics of massless adjoint Majorana fermions is the same
as that for $N_f = N_c$ flavours of massless fundamental Dirac fermions.

Gross {\rm et al.} \cite{gross} have emphasized that this should imply
screening of fundamental sources in the presence of dynamical
massless adjoint fermions. Such screening  obviously occurs
in the presence of dynamical fundamental fermions (unless $N_c >> N_f$)
since the flux line can break, but it is far from obvious that this
should occur in the adjoint case. Although the universality 
results of ref.\cite{kutasov2} constitute a (physicist's)  proof,
one would nevertheless like to understand the
screening behaviour of massless adjoint fermions from a more direct
and physical viewpoint. Various arguments were advanced in ref.\cite{gross}
in evidence of this conclusion, and the conclusion that screening disappears
if the adjoint fermions are given a mass, when the equivalence with 
fundamental fermions no longer holds. The most powerful of these arguments
showed that the Wilson loop obeys the appropriate area or perimeter law
according to whether the adjoint fermions are massive or massless
respectively. It was also emphasized that 
the Schwinger model \cite{schwinger}, where fractional charges can be screened
by massless integer charges, was an abelian prototype of this
behaviour. 

In this paper, the correspondence with the Schwinger mechanism ---
a two-dimensional dynamical version of the Higgs phenomenon ---
is made more explicit. In the large-$N_c$ $QCD_2$ with 
adjoint fermions, the vacuum polarization of the gluon is 
calculated in light-cone Tamm-Dancoff formalism, showing that 
adjoint fermions screen the linear Coulomb potential between heavy 
fundamental sources
when massless but not when massive. A composite 
bosonic state transforming in the adjoint representation of global colour
symmetry is found to be responsible for this non-abelian Schwinger mechanism,
in rather direct analogy with Schwinger's massive photon. For massless
adjoint fermions, the singlet spectrum of single-particle states is 
built entirely out of these bosons. The light-cone analysis has many
similarities with that of the usual Schwinger model \cite{berg}.

The large-$N_c$ limit is used in this paper since it illustrates
particularly clearly the phenomena in question. 
There has been a large amount of recent work on the vacuum properties
at finite $N_c$, mostly for $SU(2)$, a (probably incomplete) list of which
is refs.\cite{alot}.

\section{Adjoint $QCD_2$ and Screening.}
The action for $1+1$ dimensional $SU(N_c)$ gauge theory coupled to 
Majorana fermions $\Psi$ of mass $m$ in the adjoint representation is
\be
S= \int d^2 x \Tr \ \left\{ {\rm i} \overline{\Psi} \gamma_{\a} D^{\a} \Psi
 - m \overline{\Psi} \Psi - {1 \over 4 g^2} F_{\a\b}F^{\a\b} \right\} \ .
\eq
The conventions of ref.\cite{dalley1} will largely be followed (in particular
with regard to normal-ordering).
In the light-cone 
formalism $x^+$ is treated as `time' and $x^-$ as `space'
and we use the light-cone gauge
$A_- = (A_{0} - A_{1})/\sqrt{2} = 0$. Then $A_+$ and the left-moving components
of $\Psi$ are eliminated by their constraint equations of motion to yield
a light-cone hamiltonian 
\be
P^- = \int dx^- \Tr \ \left\{ {m^2 \over 2} \psi {1 \over {\rm i} \partial_-}
\psi  + {g^2 \over 2} J^+ {1 \over ({\rm i} \partial_-)^2} J^+ 
\right\} \ . \label{ham}
\eq
The traceless hermitian fermionic matrices $\psi_{ij}$ 
are the propogating right-moving components of $\Psi$
 while the current $J^{+}_{ij} = 2 
\psi_{ik} \psi_{kj}$, with $i \in \{ 1, \cdots, N_c \}$ (at large $N_c$
we are justified in not subtracting the appropriate Traces to distinguish 
$SU(N_c)$ from $U(N_c)$). 
The only remnants of the constrained degrees of freedom
are the instantaneous propagators $1/\partial_-$ and $1/(\partial_-)^2$
of the left-movers and $A_+$ respectively, which appear in the mass term
and current term respectively in (\ref{ham}).
The Hilbert space at fixed light-cone time $x^+$ is 
constructed from the Fourier modes of $\psi$,
\be
\psi(x^-) = {1 \over \sqrt{2 \pi} }
\int_{-\infty}^{+\infty} d\p \tsi(\p) {\rm e}^{-{\rm i} \p x^-} \ ,
\eq
by applying creation operators $\tsi(\p)$, $\p<0$, to a Fock vacuum $|0>$
annihilated by $\tsi(\p)$, $\p>0$. To any finite order in a Tamm-Dancoff
truncation on the number of fields $\tsi$ in the Hilbert space,
only singlet states under the 
residual global colour transformation $U^{\d} \psi U$, $U \in SU(N_c)$, are
annihilated by $\tilde{J}^+(p^+ = 0)$ and so avoid the $1/(p^+)^2$ singularity
of (\ref{ham}) at $p^+ =0$. There are also zero modes $\tsi(0)$ which 
form representations of an $SU(N_c)$ affine Lie algebra when applied to
the Fock vacuum $|0>$.  Ignoring them is 
valid at large $N_c$ and $m=0$ for the
bosonic single-particle boundstates \cite{kutasov2}, and also likely
to be a good approximation for large $m$ since 
the endpoint of the wavefunction in momentum space is suppressed.

\begin{figure}
\centering
\BoxedEPSF{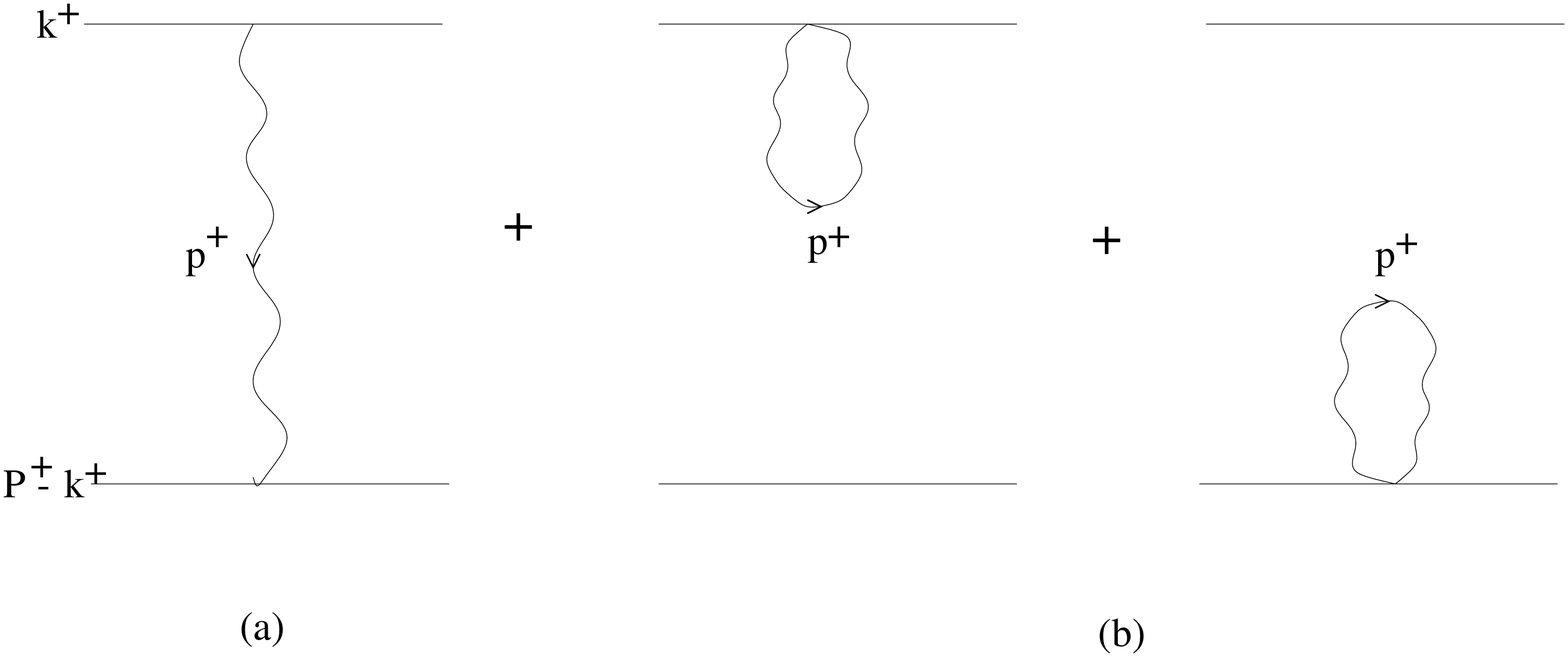 scaled 400}
\caption{Elementary processes of light-cone perturbation theory
contributing to the boundstate equation (\ref{lct}).
Wavey lines are instantaneous gluons, plain lines are
fundamental fermions. Evolution in $x^+$ to the right is understood. 
The (bare) gluons propogate in $x^-$ but not in $x^+$; they are
instantaneous. The self-energy diagrams (b), corresponding to the
last term in eq.(\ref{lct}), effect the principal-value nature of the
Coulomb exhange process.
\label{fig1}}
\end{figure}

One may easily show that the 
instantaneous gluon propogator $1 / (\partial_-)^2 $ corresponding
to $A_+$ gives rise
to a linear potential between two heavy coloured sources. The light-cone
Schrodinger equation for the wavefunction
$\phi(\k,P^+ -\k)$ of a 
pair of fundamental fermions of mass $m_F$ and momentum
$k^+$ and $P^+ -k^+$ is 't Hooft's equation \cite{hoof}
\begin{eqnarray}
2 P^- \phi(\k,P^+ -\k) & = &
m_{F}^{2} \left({1 \over \k} + {1 \over P^+ -\k}\right) \phi(\k,P^+ -\k)
\nonumber \\
&& -{\g} \int_{-\k}^{P^+-\k} {d\p \over (\p)^2} 
[\phi(\k+\p,P^+ -\k-\p) - \phi(\k,P^+ -\k) ]
\label{lct}
\end{eqnarray}
where $\g=g^2 N_c /\pi$ is held fixed in the large $N_c$ limit.
The elementary processes which comprise this equation are illustrated
by the diagrams of light-cone perturbation theory in Figure~\ref{fig1}.
When $m_F$ is large the non-relativistic (equal-time) Schrodinger equation
may be derived from (\ref{lct})
by expanding in powers of velocity and $1/m_{F}$
\cite{stan}. We have
\begin{eqnarray}
\k & = & \sqrt{m_{F}^{2} + (k^1)^2} + k^1 \nonumber \\
& = & m_F + k^1 +  {(k^1)^2 \over 2 m_F} + \cdots
\end{eqnarray}
The wavefunction is peaked at $\k= P^+ /2$  and one may derive
an equation in terms of the relative equal-time momentum $q^1 \approx 2m_F
(1-2(k^+/P^+)) << m_F$ of the
pair of fermions;
\be
{(q^1)^2 \over 4 m_F} \phi(q^1) - {\g} \int_{-\infty}^{\infty}
{dp^1 \over (p^1)^2} [\phi(p^1 + q^1) - \phi(q^1) ]
= E\  \phi(q^1) \ .\label{eqt}
\eq
Here, $p^1 = 4m_F p^+ / P^+$, 
\be
\phi(q^1)= \phi \left( P^+ \left({1 \over 2} + {q^1 \over 4m_F}
\right), P^+\left( { 1 \over 2} - {q^1 \over 4m_F}\right) \right)
\eq
and $E= \sqrt{2P^+P^-}  -2m_F$ is the binding energy. In terms of the
position space wavefunction $\phi(x^1)$ this becomes 
\be
\left[ - {1 \over 4m_F} (\partial_{x^1})^2 +  V(x^1) \right] 
\phi(x^1) = E \phi(x^1) \ ,
\eq
with $V(x^1) = \pi \g |x^1|$. Note that the small $p^1$ region in (\ref{eqt}),
hence the small $\p$ region in (\ref{lct}),
governs the asymptotic behaviour of $V$ as $|x^1| \to \infty$.

\begin{figure}
\centering
\BoxedEPSF{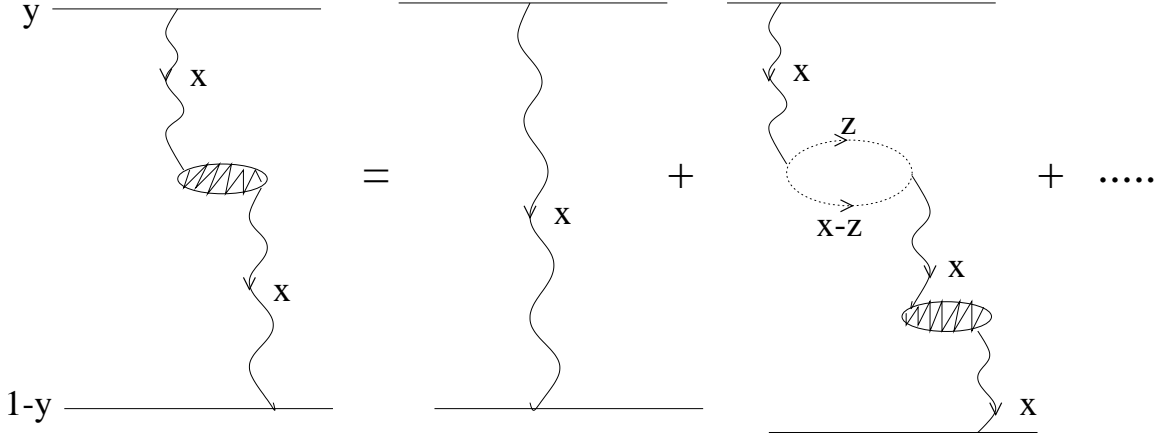 scaled 500}
\caption{Expansion in light-cone perturbation theory of the full $A_+$
propagator. Dotted lines are
adjoint fermions.\label{fig2}}
\end{figure}

If we now add adjoint fermions to the problem,
Figure~\ref{fig2} shows the expansion of the full $A_+$ propagator $G(x)$ 
in terms of the
bare instantaneous propogator $\g /x^2 $
and adjoint fermion loop corrections. We have introduced the 
momentum fractions $x=p^+/P^+$ transferred and $y = \k/P^+$ flowing 
through the external fundamental fermion, 
where $P^+$ is the 
total momentum flowing through the system.
There is also an analogous self-energy equation (see fig.1(b)). 
In light-cone Hamiltonian formalism one must remember to
divide by $(M^2 - \sum_a m_{a}^{2}/x_a)$ for every distinct
intermediate state, where $m_a$ and  $x_a$
are masses and momentum fractions of the physical 
quanta appearing in the intermediate state
and $M^2 = 2P^+ P^-$ is the invariant mass 
of the system.
The one-loop approximation to fig.~\ref{fig2} 
sums fermion bubbles;
\be
G(x)  = {\g \over x^2} + {\g \over x^2}  \cdot  \int_{0}^{x} 
dz {1 \over 
M^2 - {m_{F}^{2} \over y-x} - {m_{F}^{2} \over 1-y}- 
{m^2 \over z } -{m^2 \over x-z}}  \cdot  G(x)
\ . \label{full}
\eq
Now
\be
M^2 - {m_{F}^{2} \over y-x} - {m_{F}^{2} \over 1-y} \approx
m_F E - {(q^1)^2 \over 4} - 4 m_{F}^2 x
\eq
for large $m_F$ because $x << y$. But since $x  =  p^1/ 4 m_F $,
$p^1 \sim q^1$, and $E \sim (q^1)^2 / m_F$,
the first two terms
may be neglected compared to the third.
Evaluating the $z$-integral one then finds 
\be
G(x) = {\g \over x^2 + C(x)}
\eq
where
\begin{eqnarray}
C(x) & = & {\g \over 4 m_{F}^{2} } \left( 1 + {m^2 \over 2m_{F}^{2} x^2(
r_+ - r_-)}
\log{\left[ |r_- | \over r_+ \right]} \right) \\
r_{\pm} & = & {1 \pm \sqrt{1 + m^2/m_{F}^{2}x^2} \over 2} \ .
\end{eqnarray}
If $m \neq 0$ then $G(x \to 0) \to \g/[x^2(1+\g/6m^2)]$, so that 
the non-relativistic potential is still asymptotically linear, 
but with reduced slope. 
If $m=0$ the string tension vanishes and the
dressed gluon propogator changes to 
\be
G(x)|_{m=0} = { \g  \over  x^2  + \g/4m_{F}^{2} } 
\ . \label{prop} 
\eq
The pole in the 
gluon propogator has been cancelled
by another pole coming from the propagation of two adjoint fermions
produced with a constant wavefunction $\phi (z,x-z) = {\rm const.}$.
The corresponding non-relativistic potential is easily found to be
\be
V(x^1) = {g_N \pi \over 2} \left( 1 - {\rm e}^{-2g_N 
|x^1|} \right) \label{nonr}
\eq
At small distances the Coulomb potential $V(x^1 \to 0)
= \pi \g |x^1|$ is recovered
while at large distances the potential tends to a constant
$V(x^1 \to \infty) \to g_N \pi /2$.
Potentials similar to (\ref{nonr}) have been found from abelian \cite{gross}
and non-abelian \cite{frish} static classical solutions of the gluon
effective action, but the physical mechanism underlying it was not
apparent.

\begin{figure}
\centering
\BoxedEPSF{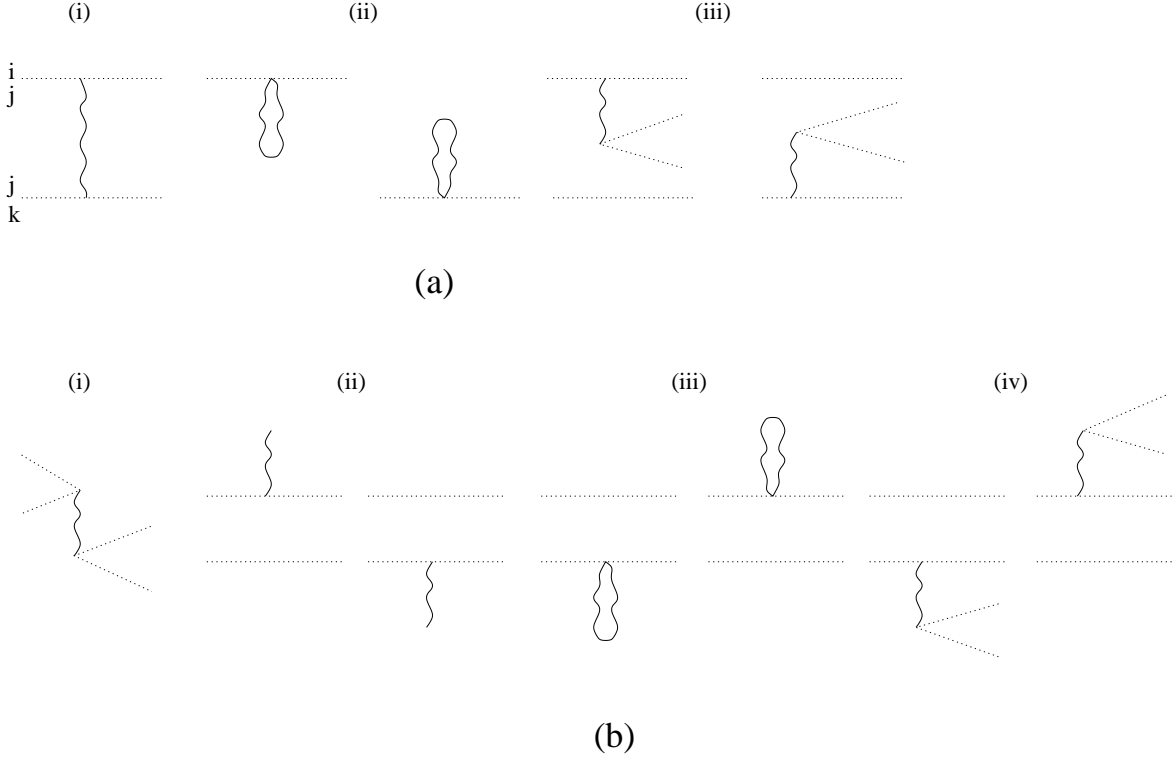 scaled 500}
\caption{This figure lists 
the elementary processes acting on $\tsi_{ij}\tsi_{jk}$
for example: (a) internal processes, which only involve the contracted
index $j$; (b) external processes, which involve one or both of 
the uncontracted indices
$i$ and $k$. Generalization to $2n$ fermions in the initial state is
straightforwardly obtained by adding the appropriate spectators, on the
outside of figs.(a)(i)(ii)(iii) and on the inside of figs.(b)(ii)(iii)(iv);
fig.(b)(i) may become either internal or external when spectators are added.
\label{fig3}}
\end{figure}

The $m=0$ result Eq.(\ref{prop}) 
is actually much more general than  1-loop 
since it includes {\em all} further diagrams 
inside the fermion loop. The reason is that all these processes cancel.
The planar topology of the large $N_c$ limit allows us to understand 
this result by writing a meaningful boundstate equation for adjoint 
states,\footnote{Here, the $w_b$ are fractions of the fraction $x$.}
\be
|\Psi>_{\rm adj} = \sum_{n=1}^{\infty} \int_{0}^{1}\prod_{b=1}^{2n} dw_{b}
{\delta(\sum_{b} w_b -1) \over N^n} \phi_{2n}( w_1, w_2, \ldots, w_{2n})
\{ \tsi(-w_1)\tsi(-w_2) \cdots \tsi(-w_{2n}) \}_{ik} |0> \ ,
\eq
provided we keep only the `internal' processes involving  contracted
colour indices (Figure~\ref{fig3}~(a)). In this case the corresponding
light-cone Schrodinger equation
$2P^+ P^- |\Psi>_{\rm adj} = M^{2}_{\rm adj} |\Psi>_{\rm adj} $
has an exact  $M_{\rm adj}^{2}=0$ two-particle solution
with constant wavefunction $\phi_{2} = {\rm const.}$,
$\phi_{2n} = 0$ for $n>1$. 
It is straightforward to verify
this on the boundstate integral equations \cite{klebanov1}
for $|\Psi>_{\rm adj}$ restricted to the processes of fig.3(a); the
non-trivial equations are
\be
M^{2}_{\rm adj}
\phi_2(w_1,1-w_1) =  
\g \int_{-w_1}^{1-w_1} {dw \over w^2} [\phi_2(w_1, 1-w_1) - 
\phi_2(w_1+ w , 1-w -w_1)] \ , \label{first}
\eq
\begin{eqnarray}
&& M^{2}_{\rm adj} \phi_4(w_1, w_2, w_3, 1-w_1 -w_2 -w_3)  =  0 \nonumber \\
&& =
{\g \over (w_2 + w_3)^2} \phi_2(w_1, 1-w_1)
-{\g \over (w_2 + w_3)^2} \phi_2(w_1+ w_2 + w_3, 1-w_1, w_2, w_3) \ .
\end{eqnarray}
The constant two-particle wavefunction remains an eigenstate if we add
the process of fig.3(b)(i), which allows it to mix with the longitudinal
component of the gluon. 
We must add
\be
\g \int_{0}^{1} dw \ \phi_2(w,1-w) 
\eq
to eq.(\ref{first}).
The mass eigenvalue is then shifted to $M^{2}_{\rm adj}
=\g$ (the Higgs mechanism).
Since other solutions of the adjoint boundstate equation are orthogonal
to $\phi_2 = {\rm const.}$, they do not contribute as intermediate
states in the vacuum polarization.
The constant wavefunction two-particle adjoint state is 
the non-abelian analogue of 
Schwinger's boson. Unlike Schwinger's massive photon,
it does not occur simply as a physical state because it is coloured.
Its definition required
us to drop `external' processes involving the uncontracted colour indices
(Figure 3(b)). In the Tamm-Dancoff approximations these involve
uncancelled infinities unless the adjoint boson is inserted inside
an overall singlet. These processes cause it to interact with other
coloured states through Coulomb exchange (fig.3(b)(ii)(iii))
and creation of further bosons (fig.3(b)(iv)).

Another way to see  why the constant wavefunction two-particle
adjoint state
plays a privileged role when $m=0$ is to note that it is 
simply the current $\tilde{J}^{+}_{ik}(-\p) |0>$ acting on the
vacuum. As emphasized in ref.\cite{kutasov2}, the hamiltonian (\ref{ham})
is expressed entirely in terms of the $SU(N_c)$
currents $J^+ = J^{+a} {\bf T}^a$
when $m=0$, which satisfy
an affine Lie algebra
\be
[\tilde{J}^{+a}(p), \tilde{J}^{+b}(k)] = k N_c \delta_{p+k,0}\delta^{ab} 
+ {\rm i} f^{abc} \tilde{J}^{+c}(p+k) \ ,
\eq
and the boundstate problem can be framed algebraicly in terms
of the bosonic basis of $\tilde{J}^+$'s rather than the fermionic
basis of $\tsi$'s.\footnote{In general at $m=0$ the basis of $\tsi$'s will
contain multi-particle combinations of the basis of $\tilde{J}^+$'s 
\cite{kutasov2}.}
Single-particle bosonic physical states  are of the form
\be
\sum_{n=2}^{\infty} 
\int_{0}^{P^+} \prod_{a=1}^{n} d\k_a
\delta(\sum_a \k_a - P^+) 
h_{n}(\k_1, \ldots, \k_n) \Tr \{\tilde{J}^+(-\k_1)
\cdots  \tilde{J}^+(-\k_n) \} |0> \ .
\eq
The light-cone boundstate integral equations for $h_n$ have been 
worked out in detail in ref.\cite{arm}, and are similar
to the corresponding ones in the fermionic basis \cite{klebanov1}.
There are also fermionic boundstates, related by a softly broken
supersymmetry \cite{kutasov1,kutasov4}.

When $m>0$, the non-abelian Schwinger boson is no longer an exact
eigenstate of the adjoint boundstate equation defined above, and many other
intermediate states contribute to the vacuum polarization in a complicated
way.
At large $m>>g^2 N$ 
we should recover the bare Coulomb amplitude.
There is evidence that confinement is lost at high temperature however
\cite{kutasov1,semenoff}, at least for large $m$.

\section{Summary.}

It has been shown how a non-abelian 
analogue of the Schwinger boson is responsible
for screening of heavy sources in large-$N_c$ $QCD_{2}$ with massless
adjoint Majorana fermions. The heavy source potential was calculated
from vacuum polarization of the gluon; every gluon propogator may
be replaced by the screened version (\ref{prop}). When the adjoint fermions
are massive, although there are additional
states contributing to the vacuum polarization, which were not included
in the calculation beyond one-loop,
the confining result found here is probably a good guide to the exact
behaviour.
Since Schwinger's work, physicists
have come up with innumerable ways to look at the Schwinger model and
no doubt many of them have an analogue in Adjoint $QCD_2$. These
investigations are left for the future, but we mention that it would be
interesting to understand how the results relate to the those found
in the bosonized formalism \cite{boson} 
and also to see explicitly what happens to gauge
invariance.

\vspace{10mm}

\noindent
{\bf Acknowledgements:} 
I thank D.Kutasov for a discussion.

\end{document}